\documentclass[twocolumn,prd,preprintnumbers,amsmath,amssymb,groupedaddresst]{revtex4}
\usepackage{graphicx}
\usepackage{color}
\newcommand{\comment}[1]{}

\newcommand{\la}{\langle}
\newcommand{\ra}{\rangle}

\usepackage[normalem]{ulem}  

\renewcommand\sout{\bgroup \color{red}\ULdepth=-.5ex \ULset}

\begin{document}

\preprint{KUNS-2526, YITP-14-84}

\title{Time evolution of gluon coherent state and its von Neumann entropy 
in heavy-ion collisions}

\author{Hideaki Iida}
\affiliation{Department of Physics, Kyoto University, Kyoto 606-8502, Japan}
\author{Teiji Kunihiro}
\affiliation{Department of Physics, Kyoto University, Kyoto 606-8502, Japan}
\author{Akira Ohnishi}
\affiliation{Yukawa Institute for Theoretical Physics, Kyoto University,
Kyoto 606-8502, Japan}
\author{Toru T. Takahashi}
\affiliation{Gumma National College of Technology, Gumma 371-8530, Japan}

\begin{abstract}
We propose a new 
prescription for evaluating a von Neumann entropy 
in the initial stage of high-energy heavy-ion collisions  
utilizing the time evolution of classical Yang-Mills (CYM) field: 
The von Neumann entropy is computed for 
the quantum coherent states constructed so as to give 
the classical gluon fields as the expectation values. 
The entropy is to be liberated 
when the complete decoherence is achieved.
As a demonstration, the time evolution of the CYM dynamics is solved
with
an initial condition which mimics the Glasma
state, though  in a non-expanding geometry;
the Glasma state
is characterized by the longitudinal color-electric and -magnetic fields
with gluon fields' fluctuations around it.
We find that the initial longitudinal fluctuations of the fields 
play essential  roles for the entropy production in two ways:
First, 
the field fluctuations at $t=0$ themselves act as a source of 
the von Neumann entropy prepared before the time evolution.
Second,
the initial fluctuations triggers field instabilities,
and hence the larger the strength of them, 
the more the entropy production at later time. 
\end{abstract}

\maketitle

\section{Introduction}
Experiments with high-energy
heavy-ion collisions (HIC) 
in Relativistic Heavy-Ion Collider (RHIC) at Brookhaven National Laboratory
and Large Hadron Collider (LHC) at CERN
revealed interesting properties
of created hot matter called the Quark-Gluon Plasma (QGP). 
One of the curious properties found in HIC experiments is 
the early thermalization;  
the thermalization time is estimated to be 
$\tau_0\simeq 1.0~{\rm fm/}c$ from 
the analysis of the data obtained 
by RHIC using relativistic hydrodynamic equations~\cite{heinz,HK2001}. 
This thermalization time is considerably smaller compared with the 
prediction by perturbative QCD~\cite{Baier:2000sb}. 
To tackle the problem, we need to understand the 
non-equilibrium QCD dynamics starting from the realistic initial
condition for HIC. 

The key issues for elucidating the possible mechanism 
 of the early thermalization 
are how and when entropy is produced in the early stage of HIC. 
Nuclear wave functions before the collision
are considered to be described by the color glass condensate 
caused by the gluon saturation~\cite{glr1981,mq1986,MV1994}, where the 
(semi-)classical approximation is  
valid~\cite{romatschke,BSSS2011,EG2013}. 
The matter created just after the collision of
 two nuclei has a specific configuration of the fields
called ``Glasma"~\cite{MV1994}, 
where color-electric and -magnetic fields 
are aligned parallel to the collision axis between two colliding nuclei. 
It is suggested that the Glasma state
induces some instabilities, i.e., Weibel and Nielsen-Olesen
 instabilities~\cite{W1959,Nielsen:1978rm}, which
may cause isotropization at early time, and
 hence hopefully thermalization \cite{Mrowczynski:1993qm,Iwazaki2008,FI2008,FII2009}.
One should, however, note that the matter which can be well described by the
dissipative hydrodynamics 
does not necessarily have isotropic particle distribution. 

A direct examination of the 
entropy production in the early stage of HIC is done
in Refs.~\cite{kunihiro,iida}, where Kolmogorov-Sina\"i (KS) entropy is given
as the sum of the positive  Lyapunov exponents calculated through solving the
time evolution of  CYM;
the positive Lyapunov exponents and thus Kolmogorov-Sina\"i (KS) entropy
 represent the mixing property of the system.
See \cite{Biro:1994bi,Matinyan:1986nw,Muller:1992iw,Biro:1994sh,Gong:1993xu,Bolte:1999th,Kunihiro:2008gv}
 for early investigations of the chaotic behavior of CYM.
In Ref.~\cite{iida}, the authors solved the equation of motion of CYM starting from Glasma-like initial conditions,
and found a large number of
positive Lyapunov exponents in the early stage 
when the initial field fluctuations are included,
and even for later times their number is a sizeable fraction 
of the total number of degrees of freedom, indicating 
a significant amount of entropy is produced by {\em classical gluon field} dynamics,
i.e., the dynamics in the early stage of HIC.
They emphasized that the field fluctuation at
the initial state is crucial for the entropy production
even when Glasma-like initial conditions are adopted.

While the early stage of HIC should be well described as a classical system, 
the system {\em is} nothing more than a quantum system $\vert G\ra$ for  which the 
gluon field $\hat{A}^a_{\mu}$ has an expectation value ${A_c}^a_{\mu}$;
$\hat{A}^a_{\mu}={A_c}^a_{\mu}+\hat{a}^a_{\mu}$ where
${A_c}^a_{\mu}=\la G\vert \hat{A}^a_{\mu}\vert G\ra$ and 
$\vert{A_c}^a_{\mu}\vert\gg\vert\hat{a}^a_{\mu}\vert$ in 
some measure.
The quantum statistical entropy is given by 
the von Neumann entropy $S\equiv -{\rm Tr}\rho\log\rho$, 
with $\rho$ being the density matrix of the system.
Our idea is that the quantum state of the gluon fields 
in the initial stage of HIC may  be well described by a coherent state~\cite{Glauber};
$\vert G\ra\simeq \vert {\rm coh}\ra$, where 
the coherent state $\vert {\rm coh}\ra$ is given solely in terms of the classical 
fields ${A_c}^a_{\mu}$; the explicit construction of $\vert {\rm coh}\ra$ 
is given in the next section.
Then the time evolution of the density matrix 
and hence the von Neumann entropy 
can be obtained from the classical dynamics.
 
However, the things do not go so easy:
Information loss of the system is indispensable for the entropy production.
Indeed, if the system stays in a pure state,
where the time evolution is given by a unitary operator, 
the entropy keeps zero value, as is well known. 
Thus the system must be or turn to a mixed state to produce the entropy. 
The conversion from a pure to mixed state or decoherence 
may be achieved through the interaction of 
the system with its environment~\cite{zurek}. 
The decoherence process can be nicely described 
in terms of the density matrix $\rho$;
the nondiagonal elements tends to vanish while the diagonal elements of $\rho$ 
becomes semi-positive definite in certain representation 
in the course of decoherence.
 
In Refs.~\cite{muller1,muller2,Fries:2008vp}, 
an estimate of the decoherence time $\tau_{\rm dec}$ of the gluon ensemble in HIC
is given in a leading-order approximation, and  it is argued
that $\tau_{\rm dec}\sim Q_s^{-1}$, 
where $Q_s$ is the saturation scale of the collision. 
They also estimate the amount of entropy production during the decoherence 
from the number of decohering gluons 
per a coherent domain~\cite{K2001}, and
found that the produced entropy is roughly one-third 
of that measured in the final hadron distribution.

\begin{figure}[t]
\begin{center}
\includegraphics[width=8cm]{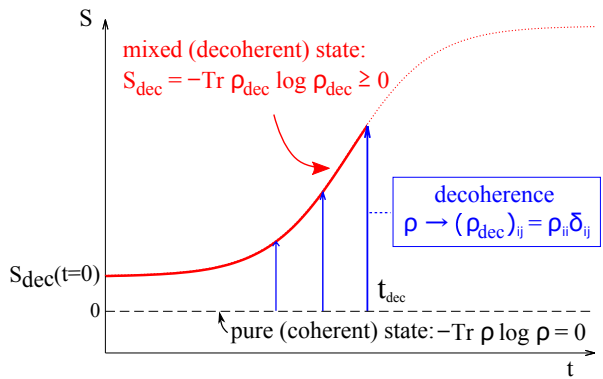}
\caption{
Schematic illustration of the possible decoherence entropy production process.
The red line shows the amount of entropy $S_{\rm dec}(t)$ to be liberated
at time $t$ provided that 
a complete decoherence occurs there, whereas 
the dashed-black line is the entropy for a pure state. 
The time evolution of the system is assumed to be described as that of a coherent state
until a time $t_{\rm dec}$ when decoherence takes place; 
then the density matrix becomes diagonal so that 
 the von Neumann entropy becomes positive. 
A possible time-dependence of entropy after $t_{\rm dec}$ is  expressed by the dotted-red line.
}
\label{pic_dec}
\end{center}
\end{figure}

In the present paper, 
we propose a new method for the calculation of the entropy 
 in the initial stage of HIC 
on the basis of a new scenario of the entropy production by decoherence.  
As is illustrated in Fig.~\ref{pic_dec}, we assume that 
there is a stage where the gluon system 
evolves quantum-mechanically  from 
Glasma initial conditions 
before the occurrence of the decoherence at a time $t_{\rm dec}$: 
The gluon system is assumed to be described by a coherent state 
as in Refs.~\cite{muller1,muller2,Fries:2008vp}, 
and the time dependence of the expectation value of gauge fields by 
the coherent state  
is obtained from that of the classical gluon fields. 
In this study, 
we do not refer to any specific process of the decoherence in the study, and 
simply assume
that a complete decoherence occurs at $t_{\rm dec}$.
Then the density matrix 
becomes diagonal in the gluon Fock state basis $(\rho_\mathrm{dec})_{ij}=\rho_{ii}\delta_{ij}$,
whose diagonal matrix elements are given 
in terms of the time-dependent coherent state, 
and thereby we can calculate
 the von Neumann entropy of the system from the diagonalized decoherent density matrix 
$\rho_\mathrm{dec}$ 
constructed from CYM dynamics, its best quantum analogue (the time-dependent coherent states), 
and the decoherence assumption.
Notice that  the entropy obtained through a completed decoherence 
gives a maximum one produced by a
possible decoherence of a coherent state. 
This kind of entropy is referred to as 
``decoherence entropy", because the 
diagonalization of the density matrix is caused 
by decoherence of a quantum state \cite{zurek}.

As is mentioned above, the chaotic behavior of gluon fields 
is related to the production of KS entropy, 
and the fluctuation of the initial state is crucial for the production \cite{kunihiro,iida}. 
As we will see, 
longitudinal fluctuations in the  Glasma initial
conditions make the von Neumann entropy 
much larger, as is the case for Kolmogorov-Sina\"i entropy. 
This indicates that the initial fluctuation of classical chromo-electric and magnetic fields is crucial 
for the production of the entropy, and 
the production of the entropy would explain the 
early thermalization. 

The paper is organized as follows. 
In Sec.~IIA, the formalism of calculations of time dependence 
of gauge fields and electric fields 
is given  in the non-compact formalism\cite{kunihiro,iida}. 
In Sec.~IIB, basic ingredients for coherent states are presented. 
The von Neumann entropy defined by a density matrix is introduced here. 
The setup of initial conditions in a
non-expanding geometry which mimics the 
Glasma initial condition is illustrated in Sec.~IIC. 
In Sec.~III, numerical calculations of the time dependence of the entropy with various longitudinal fluctuations are shown. 
The time dependence of squared electric fields are also shown to see the 
isotropization of the system. Section~IV is devoted to summarize and conclude the present work.

\section{Decoherence entropy from classical Yang-Mills dynamics}
\subsection{Formulation of classical Yang-Mills dynamics}

First, we briefly give our formulation of calculational scheme of the CYM dynamics. 
The details can be found in Ref.\cite{kunihiro,iida}. 
In pure Yang-Mills theory in the temporal gauge $A^a_0=0$, 
the Hamiltonian in the non-compact $(A,\, E)$ scheme 
takes the following form on the lattice
\begin{align}
H=&\frac{1}{2}\sum_{x,a,i}E_i^a(x)^2 + \frac{1}{4}\sum_{x,a,i,j}F_{ij}^a(x)^2 
\ ,\label{eq1} \\
F_{ij}^a(x) &= \partial_i A_j^a (x)-\partial_j A_i^a (x)
+ \sum_{b,c} f^{abc}A^b_i(x)A^c_j(x)
\ , \label{eq2}
\end{align}
where $\partial_i$ is the central difference operator in the $i$-direction,
$\partial_i A_j(x) \equiv \{ A_j(x+\hat{i})-A_j (x-\hat{i}) \}/2$, and
$f^{abc}$ is the structure constant. 
In this study, we deal with SU(2) gauge theory, 
i.e., $f^{abc}=\epsilon^{abc}$, where $\epsilon^{abc}$ is the Levi-Civita tensor with $\epsilon^{123}=+1$. 
Note that all the quantities in the equations are given in the lattice
unit and thus dimensionless. 
From Eqs.~\eqref{eq1} and \eqref{eq2}, we obtain the classical 
equations of motion (EOM) 
for the canonical coordinate $A_i^a(x)$ (gauge fields) and the canonical momentum $\,E_i^a(x)$ (color electric fields), 
\begin{align}
\dot{A}_i^a(x) &= E_i^a(x) \ ,\\
\dot{E}_i^a(x) &= \sum_{j} \partial_j F_{ji}^a(x)+ \sum_{b,c,j} f^{abc}A^b_j(x) F_{ji}^c(x),
\end{align}
which we solve with CGC-like initial conditions to be specified later. 
The fourth-order Runge-Kutta method is adopted to solve the EOM.  
We check that the violation of Gauss' law is not harmful for our numerical simulation in the non-compact formalism: 
 the number of positive, negative and zero Lyapunov exponents exists in equal amount 
 (for details, see the appendix of Ref.{\cite{iida}).

\subsection{CYM coherent state, decoherence, and von Neumann entropy}
Here we give the 
calculational method of a von Neumann entropy in the initial stage of HIC. 

 To begin with, as an example, we consider a one-degrees-of-freedom 
system described by the Hamiltonian $H=p^2/2m+m\omega^2q^2/2$, 
 where $q$ and $p$ are the canonical coordinate and momentum. 
 We set $\hbar=1$ in the following.
 The creation and annihilation operators are defined by 
 $a^\dagger=\sqrt{\frac{m\omega}{2}}(q-ip/m\omega)$ and $a=\sqrt{\frac{m\omega}{2}}(q+ip/m\omega)$, respectively, 
 and satisfy $[a,a^\dagger]=1$. } 
 Using the creation operator, a Fock state is expressed by 
 $|n\rangle\equiv\frac{1}{\sqrt{n!}}(a^\dagger)^n|0\rangle$, 
 and is normalized as $\langle n|n^\prime\rangle=\delta_{n,n^\prime}$. 

A coherent state $|\alpha\rangle$ is an eigenstate of an annihilation operator $a$, i.e.,   
$a|\alpha\rangle =\alpha|\alpha\rangle$, where the coherent state is normalized as $\langle \alpha |\alpha \rangle=1$. 
The eigenvalue $\alpha$ of the coherent state $|\alpha\rangle$ is written by 
the expectation value of the coordinate $q$ and the momentum $p$ as 
\begin{align}
\alpha=(m\omega\langle\alpha |q|\alpha\rangle+i\langle\alpha |p|\alpha\rangle)/\sqrt{2m\omega}.
\end{align} 
Such a coherent state 
can be realized as the superposition of Fock states, 
\begin{align}
|\alpha\rangle=\exp(-|\alpha|^2/2)
\sum_{n=0}^{\infty}\frac{\alpha^n}{\sqrt{n!}}|n\rangle.
\end{align} 
The probability of finding the $n$-particle state in the coherent state is then given by
\begin{align}
P(n)\equiv |\langle n|\alpha \rangle|^2=\exp(-|\alpha|^2)|\alpha|^{2n}/n!,
\end{align}
which is a Poisson distribution. 
The average particle number $\bar n$  is then written as 
\begin{align}
\bar n\equiv \sum_n n |\langle n |\alpha\rangle|^2=|\alpha|^2. 
\label{nbar}
\end{align}

We now proceed with the case of QCD, 
which has infinite degrees of freedom.
The coherent states of gluons have color ($a$),
direction ($i$) and momentum indices ($\vec k$), i.e., 
\begin{align}
a^{ia}_{\vec k} |\alpha^{ai}_{\vec k}\rangle=\alpha^{ia}_{\vec k} |\alpha^{ai}_{\vec k}\rangle.
\end{align}
Thus the eigenvalue $\alpha^{ai}_{\vec k}$ is the expectation value of 
an annihilation operator $a^{ia}_{\vec k}$ and is expressed by
the expectation values of gauge fields and electric fields:  
The gauge fields $A^{ai} (\vec x)$ and electric fields $E^{ai} (\vec x)$ 
are expressed in terms of creation and annihilation operators, ${a^{ai}_{\vec k}}^\dagger$ and ${a^{ai}_{\vec k}}$, as
\begin{align}
A^{ai}(\vec x)&=\frac{1}{V}\sum_{\vec k} \frac{1}{\sqrt{2\omega_{\vec k}}}
\left\{ e^{i\vec k \vec x-i\omega t} a^{ai}_{\vec k}+e^{-i\vec k \vec x+i\omega t} {a^{ai}_{\vec k}}^\dagger\right\},\nonumber \\
E^{ai}(\vec x)&=\frac{1}{V}\sum_{\vec k} \frac{-i\omega_{\vec k}}{\sqrt{2\omega_{\vec k}}}
\left\{ e^{i\vec k \vec x-i\omega t} a^{ai}_{\vec k}-e^{-i\vec k \vec x+i\omega t} {a^{ai}_{\vec k}}^\dagger\right\}. 
\end{align}
Here, $\omega_{\vec k}$ on the lattice is defined by
\begin{align}
\omega_{\vec k}&\equiv \sqrt{\sin^2 k_x+\sin^2k_y+\sin^2k_z},\\
\vec k&\equiv \left(\frac{2\pi n_x}{N_x}, \frac{2\pi n_y}{N_y}, \frac{2\pi n_z}{N_z}\right), \\
0&\le n_i\le N_i-1, n_i\in \mathbb{Z} \ \ (i=x,y,z),
\end{align}
where $N_i$ is the number of lattice points in the $i$-direction.  
The expectation values of gauge and electric fields for
coherent states are then expressed as
\begin{align}
\langle A^{ai}(\vec x) \rangle &=\frac{1}{V}\sum_{\vec k} \frac{1}{\sqrt{2\omega_{\vec k}}}
\left\{ e^{i\vec k \vec x-i\omega t} \langle a^{ai}_{\vec k}\rangle +e^{-i\vec k \vec x+i\omega t} \langle {a^{ai}_{\vec k}}^\dagger\rangle \right\},\nonumber \\
\langle E^{ai}(\vec x) \rangle &=\frac{1}{V}\sum_{\vec k} \frac{-i\omega_{\vec k}}{\sqrt{2\omega_{\vec k}}}
\left\{ e^{i\vec k \vec x-i\omega t} \langle a^{ai}_{\vec k}\rangle -e^{-i\vec k \vec x+i\omega t} \langle {a^{ai}_{\vec k}}^\dagger\rangle \right\}. 
\end{align}
In the momentum space, these expectation values are written as
\begin{align}
\langle A^{ai}(\vec k)\rangle &= (\langle a^{ai}_{\vec k} \rangle +\langle {a^{ai}_{\vec k}}^\dagger\rangle)/\sqrt{2\omega_{\vec k}},\\
\langle E^{ai}(\vec k)\rangle &= (-i\omega_{\vec k})(\langle a^{ai}_{\vec k} \rangle -\langle {a^{ai}_{\vec k}}^\dagger\rangle)/\sqrt{2\omega_{\vec k}}.
\end{align}
From the above equations, the eigenvalue $\alpha^{ai}_{\vec k}$, i.e., the expectation value of the annihilation operator, is expressed as
\begin{align}
\alpha^{ai}_{\vec k}=
\langle a^{ai}_{\vec k}\rangle&=\frac{1}{\sqrt{2\omega_{\vec k}}}(\omega_{\vec k}\langle A^{ai}(\vec k)\rangle
+i\langle E^{ai}(\vec k)\rangle ).
\end{align}
The corresponding expansion of coherent states $|\alpha^{ai}_{\vec k}\rangle$ by Fock states $|n,a,i,\vec k\rangle$ is written as
\begin{align}
|\alpha^{ai}_{\vec k}\rangle=\exp(-|\alpha^{ai}_{\vec k}|^2/2)
\sum_{n=0}^{\infty}\frac{\alpha^n}{\sqrt{n!}}|n,a,i,\vec k\rangle.
\end{align} 
The probability of finding the $n$-particle state, $P^{ai}_{\vec k}(n)$ in the coherent state is given by
\begin{align}
P^{ai}_{\vec k}(n)\equiv |\langle n,a,i,\vec k|\alpha \rangle|^2=\exp(-|\alpha^{ai}_{\vec k}|^2)|\alpha^{ai}_{\vec k}|^{2n}/n!.
\end{align}
Also, the average particle number $\bar n^{ai}_{\vec k}$ is  
\begin{align}
\bar n^{ai}_{\vec k}\equiv \sum_n n |\langle n, a, i, \vec k |\alpha^{ai}_{\vec k}
\rangle|^2=|\alpha^{ai}_{\vec k}|^2. 
\label{nbar}
\end{align}

Because $\langle A^{ai}(\vec k)\rangle$ and $\langle E^{ai}(\vec k) \rangle$ 
are the expectation values for coherent states, which are the best quantum analogue of 
classical observables, 
{\it we substitute 
the classical fields $A^{ai}(\vec k)$ and $E^{ai}(\vec k)$
obtained by solving CYM equation for
$\langle A^{ai}(\vec k)\rangle$ and 
$\langle E^{ai}(\vec k) \rangle$}, 
which enables us to compute 
the time dependence of $\alpha^{ai}_{\vec k}$ with CYM.

From the coherent states of gluon fields, 
we can evaluate the density matrix $\rho$ and the 
von Neumann entropy after decoherence as follows. 
The gluon state is assumed to be a direct product of gluon coherent state in the calculation,
\begin{align*}
\rho=\prod_{a, i, \vec k}|\alpha^{ai}_{\vec k}\rangle \langle\alpha^{ai}_{\vec k}|. 
\end{align*}
The density matrix for a certain $\vec{k}, a$ and $i$ can be written as 
\begin{align}
\rho^{ai\vec{k}}_{m,n}
=\langle m,ai\vec{k}|\alpha^{ai}_{\vec{k}}\rangle
\langle\alpha^{ai}_{\vec{k}}|n, ai\vec{k}\rangle,
\end{align}
where $|m, ai\vec{k}\rangle$ is a Fock state 
with particle number $m$, color $a$, direction $i$ and momentum $\vec{k}$.  
A von Neumann entropy is defined as
\begin{align}
S=-\rm{Tr}(\rho\ln\rho)
\ .
\end{align}
For a pure state,
the entropy is zero,
because the pure state condition $\rho^2=\rho$ results in 
the eigenvalues of $\rho$ of $0$ or $1$. 
As is mentioned in Introduction, 
 we assume a complete decoherence at a time, and then 
 the density matrix becomes diagonal in the gluon Fock state basis  
$|n,ai\vec{k}\rangle$ as
\begin{align}
(\rho_\mathrm{dec})^{ai\vec{k}}_{m,n}
=
\rho^{ai\vec{k}}_{nn}\delta_{mn}
=P^{ai}_{\vec{k}}(n)\delta_{mn} \ ,
\end{align} 
and the von Neumann entropy becomes nonzero, 
\begin{align}
S_\mathrm{dec}
&=-\mathrm{Tr}\left(\rho_\mathrm{dec} \ln \rho_\mathrm{dec}\right)
\nonumber\\
&=-\sum_{n,a,i,\vec k} 
P^{ai}_{\vec k}(n)\ln P^{ai}_{\vec k}(n) \geq 0.
\label{Eq:Sdec}
\end{align}
This entropy is referred to as 
``decoherence entropy", as mentioned in Introduction.

\subsection{Initial conditions with coherent background fields}
One of the realistic initial conditions of high-energy heavy-ion collisions is 
so-called Glasma~\cite{MV1994,romatschke,fukushima,fukushima2}, 
where both the color-electric and -magnetic fields 
are parallel to the collision axis. 
In this study, we adopt a Glasma-like initial condition (GIC) 
in the non-expanding geometry, which is explained below. 

The Glasma initial condition 
describes the gauge fields at proper time $\tau\equiv \sqrt{t^2-z^2}=0^+$,
when two nuclei move along $z$ axis almost at the speed of light and collide 
at $t=0$ and $z=0$. 
The gluons having large Bjorken $x$ and quarks are assumed to 
be color sources for low $x$ gluons in the initial condition. 
The model in which the color-source distribution is 
Gaussian is called McLerran-Venugopalan model \cite{MV1994}. 

We prepare an initial condition which mimics the MV model 
in the non-expanding geometry: 
The color sources which obey Gaussian distribution are set and then
electric fields and magnetic fields aligned in the collision axis 
are generated from the sources. 
The small fluctuations in the $z$ direction is also added on the above fields 
as in Ref.~\cite{romatschke}. 

The procedure to generate color-electric and -magnetic fields is given as follows:
\begin{itemize}
\item Generate the Gaussian random color sources for a target nucleus 
$\rho^{(t)}$ and a projectile $\rho^{(p)}$ which satisfies\\
$\langle \rho^{(t)a}({\bf x_\perp}) \rho^{(t)b}({\bf y_\perp}) \rangle
=g^4\mu^2\delta^{ab}
\delta^{(2)}({\bf x_\perp}-{\bf y_\perp})$ \\
and \\
$\langle \rho^{(p)a}({\bf x_\perp}) \rho^{(p)b}({\bf y_\perp}) \rangle
=g^4\mu^2\delta^{ab}
\delta^{(2)}({\bf x_\perp}-{\bf y_\perp})$, 
 where ${\bf x}_\perp\equiv (x,y)$ and 
 $a, b$ are the color indices.   
 
\item Solving the Poisson equations: \\
$-{\bf \partial}^2_\perp\Lambda^{(t)}({\bf x_\perp})=\rho^{(t)}({\bf x_\perp})$, \\ $-{\bf \partial}^2_\perp\Lambda^{(p)}({\bf x_\perp})=\rho^{(p)} ({\bf x_\perp})$
\item Wilson lines are calculated as\\
$V^\dagger({\bf x_\perp})=e^{i\Lambda^{(t)}({\bf x_\perp})}$, \ 
$W^\dagger({\bf x_\perp})=e^{i\Lambda^{(p)}({\bf x_\perp})}$.
\item Gauge fields are given by $\alpha_i^{(t)}=iV\partial_i V^\dagger$ 
 ($i=x,y$) and 
$\alpha_i^{(p)}=iW\partial_i W^\dagger$. 
\item 
From $\alpha_i^{(t)}$ and $\alpha_i^{(p)}$, 
gauge fields are constructed as 
\begin{align}
& A^i=\alpha_i^{(t)} + \alpha_i^{(p)}, A^z=0.
\label{Eq:MVA}
\end{align}
Electric fields and magnetic fields are then given by
\begin{align}
& E^i=0, E^z 
=i\sum_i\left(\left[\alpha_i^{(t)},\alpha_i^{(p)}\right]\right),
\label{Eq:MVE}\\
& B^i=0, B^z=i\left(\left[\alpha_1^{(t)},\alpha_2^{(p)}\right]+\left[\alpha_1^{(p)},\alpha_2^{(t)}\right]\right).
\label{Eq:MVB}
\end{align}
\end{itemize}
The initial condition gives electric fields and magnetic fields 
parallel to the $z$ direction and have fluctuations in the $x$ and $y$ directions. 

We also include fluctuations in the $z$ direction as follows~\cite{romatschke}:
\begin{itemize}
\item Generate the Gaussian random fluctuations of electric fields, 
$\delta \bar E_i({\bf x_{\perp}})$, 
 which satisfies
\begin{align}
\langle \delta\bar E_i({\bf x_{\perp}})\delta\bar E_j({\bf y_{\perp}})\rangle
=\delta_{ij}\delta({\bf x_{\perp}}-{\bf y_{\perp}})
\end{align}
 and also fluctuations in $z$ direction $F(z)$ satisfying
\begin{align}
g^2\mu
 \langle F(z)F(z^\prime)\rangle=\Delta^2\delta(z-z^\prime).
 \label{etaflc}
\end{align}
\item $E^i({\bf x_{\perp}},z)$ and 
$E^z({\bf x_{\perp}},z)$ 
are composed from these fluctuations as
\begin{align}
E^i({\bf x_{\perp}},z)&=\delta E_i({\bf x_{\perp}},z)
\ ,\label{Eq:GICEi}\\
E^z({\bf x_{\perp}},z)
&=i\sum_i\left(\left[\alpha_i^{(t)},\alpha_i^{(p)}\right]\right)
+\delta E_z({\bf x_{\perp}},z)\nonumber\\
&\equiv E^{(0)}_z
+\delta E_z({\bf x_{\perp}},z)
\ ,\label{Eeta}
\end{align}
where
\begin{align}
&\delta E_i({\bf x_{\perp}},z)=\partial_z F(z) \delta\bar E_i({\bf x_{\perp}}),\\
&\delta E_z({\bf x_{\perp}},z)=-F(z)D_i \delta\bar E_i({\bf x_{\perp}})\ ,
\end{align}
and $E^{(0)}_z$ is the electric field without fluctuations
shown in Eq.~\eqref{Eq:MVE} 
\end{itemize}
GIC shown in Eqs.~\eqref{Eq:MVA}, \eqref{Eq:GICEi} and \eqref{Eeta}
with $(t, \mathbf{x}_\perp, z)$ coordinates
are given in almost the same way as those in the original 
formulation,
which is given in an expanding geometry
with $(\tau, \mathbf{x}_\perp, \eta)$ 
coordinates~\cite{MV1994,romatschke,fukushima,fukushima2},
by replacing $(\tau, \eta)$ with $(t,z)$.
Since we replace $\delta(\eta-\eta')$ in an expanding geometry
with $\delta(z-z')$ in a non-expanding geometry in Eq.~\eqref{etaflc}, 
we have introduced a scale factor $g^2\mu$ in Eq.~\eqref{etaflc}
in order to compensate the dimension coming from $\delta(z-z')$.
It should be noted that the present initial condition does not have
the connection with the color glass condensate
wave function of the colliding nuclei. 
We remark that the construction of electric fields is consistent 
with Gauss' law, $D_iE^i+D_zE^z=0$.  

\begin{center}
\begin{table}[t]
\caption{
Parameter set for the Glasma initial conditions.  
$\Delta$ denotes the amplitude of longitudinal fluctuations, $\epsilon$ is energy density 
written in $(g^2\mu)^4$ and GeV/fm$^3$, and  
``ratio of $E^2$'' is defined by $\sum_{\vec x, a}\delta E_z^a(\vec x)^2/\sum_{\vec x,a} {E_z^{(0)a}}(\vec x)^2$, where $\delta E_z^a(\vec x)$ and ${E_z^{(0)a}}(\vec x)$ are 
defined in Eq.\eqref{Eeta}. The physical scale of energy density is determined by regarding 
the transverse area of the lattice as a section of an Au nucleus (see the last paragraph of Sec.III).}
\begin{tabular}{ccccc}
\hline
\hline
volume \ \ & $\Delta$ \ \ &$\epsilon (g^2\mu)^4$ &$\epsilon($GeV/fm$^3$)& ratio of $E^2$\\
\hline
$20^3$ &0 & $0.1711$ & 0.3476 & 0\\
$20^3$ & $1.0\times 10^{-3}$ &$0.1711$ & 0.3476& $3.278\times 10^{-5}$\\
$20^3$ & $5.0\times 10^{-3}$ &$0.1712$ & 0.3478& $8.194\times 10^{-4}$\\
$20^3$ & $1.0\times 10^{-2}$ &$0.1714$ & 0.3482& $3.278\times 10^{-3}$\\
$20^3$ & $5.0\times 10^{-2}$ &$0.1814$ & 0.3685& $8.194\times 10^{-2}$\\
$20^3$ & $1.0\times 10^{-1}$ &$0.21$ & 0.4266& $3.278\times 10^{-1}$\\
\hline
\hline
\end{tabular}
\label{tab1}
\end{table}%
\end{center}

\section{Decoherence entropy from Glasma initial condition}

\begin{figure}[t]
\begin{center}
\includegraphics[width=8cm]{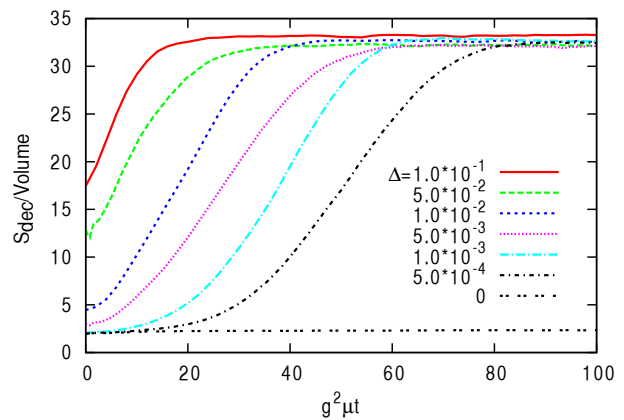}
\caption{Time dependence of density of the entropy, $S_{\rm dec}/ {\rm Volume}$, 
from Glasma initial condition 
with various fluctuations $\Delta$ in $z$ direction 
shown in Eq.\eqref{etaflc}.
}
\label{Fig:Sdec}
\end{center}
\end{figure}

In this section, we show the numerical results of the entropy evaluated with GIC.
Note that we set $g^2\mu =1$
in the numerical calculation. 
In Table \ref{tab1}, parameters used in the initial condition are summarized; volume, amplitudes of longitudinal fluctuations, and  
initial energy densities. We also show
the squared fluctuation-background electric field ratio defined by 
$\sum_{\vec x, a}\delta E_z^a(\vec x)^2
/\sum_{\vec x,a} {E_z^{(0)a}}(\vec x)^2$,
where $\delta E_z^a(\vec x)$ and ${E_z^{(0)a}}(\vec x)$ 
are defined in Eq.~\eqref{Eeta}.

Figure \ref{Fig:Sdec} shows the time dependence of the entropy density, 
$S_{\rm dec}/{\rm Volume}$, from GIC 
with fluctuations in the $z$ direction. 
The amplitude of longitudinal fluctuations, $\Delta$ in Eq.\eqref{etaflc}, 
varies from $0$ to $1.0\times 10^{-1}$. The lines correspond to 
different amplitudes of longitudinal fluctuations, $\Delta$.
We will mention the choice of the amplitudes $\Delta$ later.  
With $\Delta=0$, the entropy is almost constant:
$S_{\rm dec}/{\rm Volume}\simeq 2.3$ for our calculation time, $g^2\mu t<100$. 
On the other hand, with finite $\Delta$, $S_{\rm dec}/{\rm Volume}$ starts to increase at a certain time. 
For example, with $\Delta=5.0\times 10^{-4}$, $S_{\rm dec}/{\rm Volume}$
is almost the same as that without fluctuation for $g^2\mu t<18$, and 
it starts to increase after that and saturates around $g^2\mu t=80$. 
The onset time of entropy increase becomes earlier 
as the fluctuation $\Delta$ becomes larger. 
For $\Delta > 5.0\times 10^{-3}$, $S_{\rm dec}/{\rm Volume}$ 
at the initial time is already larger 
than that without fluctuations. 
Thus, we find the two kinds of origin of the entropy. 
One is the non-linear dynamics of CYM,
which is triggered by small fluctuations after a certain time. 
The other is the intrinsic entropy which is potentially prepared in the initial state (at $t=0$).

We here discuss the relation between time dependence of the entropy 
and the initial fluctuations.  
In the previous paper~\cite{kunihiro,iida},
chaotic behavior of CYM and the entropy production are studied. 
They find that the instability and chaotic behavior of CYM are the origin of the 
Kolmogorov-Sina\"i (KS) entropy,
which is obtained by the sum of positive Lyapunov exponents 
and shows the entropy production rate. 
On the other hand, the entropy itself is studied here. 
The numerical results obtained here indicate that 
the production of the decoherence 
entropy would be also related to instabilities and chaoticity of the system. 
Without longitudinal fluctuations, i.e., the amplitude $\Delta=0$, 
no long-standing instability exists. 
While the initial Glasma configuration has a potential instability
in the longitudinal fluctuation direction leading to macroscopic 
number of the Kolmogorov-Sina\"i entropy obtained as the sum of
positive Lyapunov exponents, actual entropy production would not
be realized because of the translational invariance in $z$ direction. 
The number of excited modes is that of two-dimensional lattice, 
and dimensionally reduced entropy from two-dimensional chaos
is realized. 
With finite $\Delta$, potential instability represented as positive Lyapunov
exponents is manifested, and full 3-dimensional entropy production
is realized. The number of degrees of freedom is proportional
to the lattice sites, and the entropy is roughly about
$N_z$ times larger than that in the $\Delta=0$ case,
where $N_z$ is the lattice size in $z$ direction 
and $N_z=20$ in the present calculation. 
Our results show that the entropy with finite $\Delta$ is 
around 16 times larger than that with $\Delta=0$, which is 80\%
of the expected ratio. 
This small reduction is understood as the difference of the 
distribution of the average particle number $\bar{n}^{ai}_{\vec k}$. 
With $\Delta=0$, energy is concentrated in smaller 
number of modes, and we find $\bar{n}$ is larger in 
excited modes, $\alpha \not=0$, then the entropy 
per excited mode is slightly larger with $\Delta=0$.


\begin{figure}[t]
\begin{center}
\includegraphics{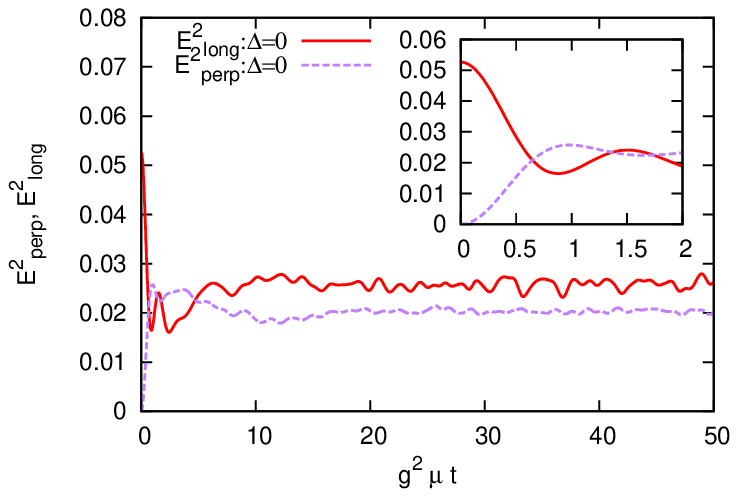}
\includegraphics{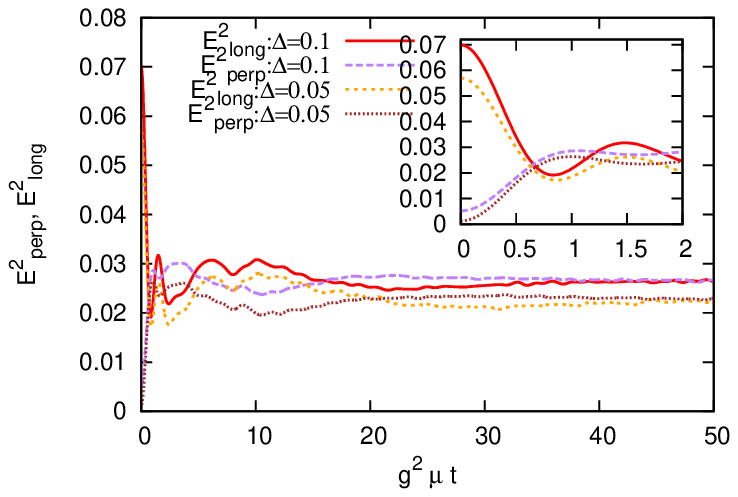}
\caption{(Upper figure): time evolution of $E^2_{\rm perp}$ and $E^2_{\rm long}$ from the Glasma initial conditions without fluctuations. (Lower figure): that with fluctuations 
($\Delta=0.1$ and $0.05$). The figures in the right-top side of the two figures are the enlarged views of each figure for $t=0$ to 2$(g^2\mu)^{-1}$ }
\label{Fig:iso}
\end{center}
\end{figure}
Next, we discuss isotropization of the system. 
In general, the isotropization time depends on physical observables. 
We here examine 
squared electric fields in the perpendicular and the longitudinal directions, 
$E^2_{\rm perp}=((E^{xa})^2+(E^{ya})^2)/2$ 
and $E^2_{\rm long}=(E^{za})^2$. 
The upper panel of Fig.\ref{Fig:iso} shows 
$E^2_{\rm perp}$ and $E^2_{\rm long}$ computed without fluctuations in the initial fields' configuration.
Initially, there is finite $E^2_{\rm long}$ and no $E^2_{\rm perp}$. 
As time goes by, $E^2_{\rm perp}$ grows and $E^2_{\rm long}$ become small. 
However, they do not coincide for $g^2\mu t<50$ as seen from the figure. 
The lower panel of Fig.\ref{Fig:iso} shows 
$E^2_{\rm perp}$ and $E^2_{\rm long}$ estimated with initial longitudinal fluctuations, 
where the amplitudes of the fluctuation, $\Delta$, are 0.1 and 0.05. 
While for $\Delta=0.1$, the initial behavior of those are similar 
to that without fluctuations, 
$E^2_{\rm perp}$ and $E^2_{\rm long}$ almost coincides after $g^2\mu t\simeq 35$, 
which indicates isotropization of the system. 
For $\Delta=0.05$, the basic behavior is the same as that for $\Delta=0.1$, 
but the isotropization time is a bit longer than that 
with $\Delta=0.1$: 
$g^2\mu t\simeq 40$ for this amplitude, 
which indicates the stronger fluctuations cause  
faster isotropization. 
We note that the ``isotropization'' here is determined 
by that of electric fields, 
and there might be some quantities that is not isotropized 
after the isotropization of electric fields. 

We briefly comment on the choice of the range of fluctuations,
$0<\Delta < 1.0\times 10^{-1}$. 
Since the longitudinal fluctuation strength is of order $g^2$ 
compared to the background field in the perturbative regime, longitudinal fluctuations 
should not be large. 
As shown in Table~\ref{tab1},
the squared fluctuation-background electric field ratio, 
$\sum_{\vec x, a}\delta E_z^a(\vec x)^2/\sum_{\vec x,a}
{E_z^{(0)a}}(\vec x)^2$, is less than $1/3\times 10^{-1}$. The fluctuations with $\Delta \leq 0.1$ are 
judged to be small compared with the background field 
and thus these fluctuations are acceptable. 

\begin{figure}[t]
\begin{center}
\includegraphics[width=8cm]{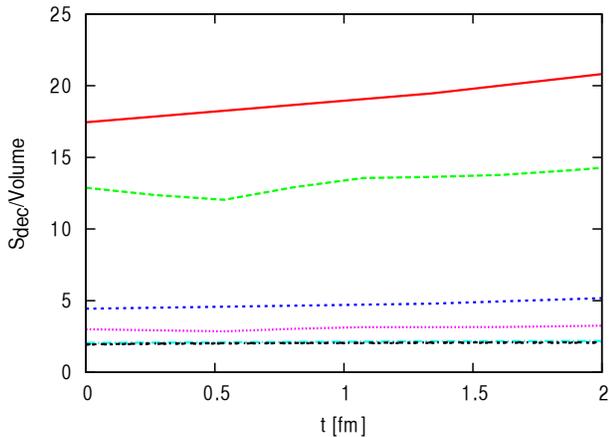}
\caption{The same figure of Fig.\ref{Fig:Sdec} with a physical scale in the horizontal axis 
for t=0 to 2fm. The lines correspond to different amplitudes of longitudinal fluctuations, $\Delta$, and 
the correspondence of line types with $\Delta$ is the same as that of Fig.\ref{Fig:Sdec}.
Note that three lines for $\Delta=0, 5.0\times 10^{-4}$ and $1.0 \times 10^{-3}$ are 
almost degenerate in the time interval. }
\label{Fig:Sdecphys}
\end{center}
\end{figure}
Rough estimation of the physical scale can be done by regarding the 
transverse area of the lattice as a section of an Au nucleus. 
This enables us to determine the physical length of the lattice, $L$, as $L^2\simeq \pi R_{\rm Au}^2$ with the radius of an Au nucleus,  
$R_{\rm Au}\simeq 7{\rm fm}$.  
The resulting physical length is $L\simeq 12.4{\rm fm}$. 
From the present lattice extent $g^2\mu L=20$,
we obtain $g^2\mu \simeq 0.32{\rm GeV}\simeq (0.63{\rm fm})^{-1}$.
Figure \ref{Fig:Sdecphys} is the same as Fig.\ref{Fig:Sdec} but 
with the horizontal axis from $t$=0 to 2fm.
From the figure,
we see that the entropy is roughly constant in the time interval, 
which implies the importance of  
the potential entropy possessed  
in the initial state and the decoherence time at which the 
decoherence entropy is realized 
 for the understanding of early thermalization.
We should note here that
the determination of the physical scale has large uncertainty and 
we also do not know how large the actual longitudinal fluctuation is. 
Further study is necessary for the deeper understanding 
of the entropy production. 

\section{Summary and conclusion}
We have studied entropy production in the initial stage of HIC. 
We have proposed a method to compute (quantum mechanical) von Neumann entropy 
utilizing the dynamics of classical Yang-Mills(CYM) fields
under the following two assumptions:
One is that the  gluon state in the initial stage 
is well represented by a quantum coherent state, and the time evolution of 
the classical gluon fields as given by the expectation values of 
the coherent state is provided by the CYM dynamics. 
The other is that the density matrix becomes fully diagonal, i.e., 
complete decoherence occurs at a time $t_{\rm dec}$, 
which gives the upper limit of the entropy produced through decoherence.
The von Neumann entropy evaluated here is referred 
to as the decoherence entropy. 
We do not specify 
the mechanism of the decoherence here.
We have adopted an initial 
condition, Glasma-like initial condition (GIC), 
which mimics the Glasma 
in the non-expanding geometry to calculate the time dependence of the entropy. 
The longitudinal fluctuation is imposed to the initial condition and 
we see the dependence of the entropy production 
on the longitudinal fluctuations. 

In GIC with small longitudinal fluctuations, 
the entropy starts to increase at a certain time 
and is enhanced compared with that without longitudinal fluctuations. 
The onset time of the entropy increase  
depends on the ratio of the fluctuations  
to background color-electric and magnetic fields. 
Thus, the longitudinal fluctuations in the initial condition 
are quite important for entropy production.  
On the other hand, with large longitudinal fluctuations, 
the decoherence entropy is already large at $t=0$. 
We find two kinds of origin of the entropy: 
One is the non-linear dynamics of CYM triggered by small fluctuations 
after a certain time,  
and the other is the intrinsic entropy potentially kept in the initial state. 

We have discussed the relation between the production 
of the decoherence entropy and the instability and chaoticity of the system. 
Without the longitudinal fluctuations, 
the instability in the longitudinal direction is not realized due to 
translational invariance in longitudinal direction and 
the entropy production is dimensionally reduced. 
On the other hand, 
a full 3-dimensional entropy production is 
realized by the longitudinal fluctuations 
due to the realization of instability and chaoticity in the longitudinal direction. Our numerical simulations support the scenario. 


%
Isotropization has also been investigated via squared electric fields. 
We find that isotropization of the system in GIC is also intimately
related to the fluctuation of electric and magnetic fields in 
the initial stage. 
We have also estimated the physical scale and find that 
the initial entropy kept in the coherent state and the decoherence time 
would be important for the understanding of the early thermalization. 
We should note that we still have uncertainty in the
determination of the physical scale and the magnitude of longitudinal
fluctuations, which should be resolved before drawing phenomenological conclusions.

In the present study, we neglect the expansion of
a Glasma along the collision axis.  
The effect of expansion would make the thermalization slower. 
Therefore, to include the effect in the calculation is an 
important future task to understand 
the entropy production and the thermalization in HIC.

{\it Acknowledgment.} 
We are grateful to Berndt M\"uller for fruitful discussions and making us aware of Refs.\cite{muller1,muller2,Fries:2008vp} after the initiation of the present work. 
The calculations were performed mainly by using the NEC-SX9 at
Osaka University. 
This work was supported in part by
 Grant-in-Aid for Scientific Research from
 the Japan Society for the Promotion of Science (JSPS)
 and the Ministry of Education, Culture, Sports, Science and Technology
 of Japan (MEXT)
 (Nos.
 20540265 
 Innovative Areas (No. 2004: 23105713, and No. 2404: 24105001, 24105008), 
 23340067, 
 24340054, 
 24540271
),
 by the Yukawa International Program for Quark-Hadron Sciences,
 by a Grant-in-Aid for the global COE program
``The Next Generation of Physics, Spun from Universality and Emergence''
from MEXT. 
T. K. is supported by  the Core Stage Back UP program 
in Kyoto University.


\end{document}